Room temperature study of a strain-induced electronic superstructure on a magnetite (111) surface.

N.Berdunov, S.Murphy, G.Mariotto, I.V.Shvets
SFI Nanoscience laboratory, Physics Department, Trinity College, Dublin 2, Ireland

Abstract

A magnetite ($Fe_3O_4$) single crystal (111) surface has been studied at various oxygen-iron surface stoichiometries. The stoichiometry was modified by controlling the *in-situ* sample anneal conditions. We have found the conditions that lead to the formation of an oxygen-rich surface that forms a quasi-hexagonal superstructure with a 42Å periodicity. The superstructure is highly regular and was observed by both LEED and STM. The superstructure consists of three regions, two of which have identical atomic scale structures with a periodicity of 2.8 Å, and a third having a periodicity that is about 10% larger (3.1 Å). The subtle difference in the atomic periodicities between the three areas results from the modulation of intrinsic strain developed along the surface. The superstructure results from electronic effects rather than being a mosaic of different iron oxide terminations. The onset of the superstructure is sensitive to the surface stoichiometry. From our results we could estimate the critical density of defects leading to the disappearance of the superstructure. We have modelled the experimental results and calculated the electron density using a DFT algorithm. The model clearly shows the development of strain along the surface.

1. INTRODUCTION

Magnetite is a well-known example of a material that undergoes a metal-insulator transition, one of the most important phenomena in solid-state physics. Magnetite has the inverse spinel crystal structure based on an fcc lattice of oxygen ($O^{2-}$) anions, containing Fe cations in tetrahedrally (so-called A sites) and octahedrally (B sites) coordinated interstices (Fig.1). The A-sites are occupied by $Fe^{3+}$ ions and the B-sites by $Fe^{2+}$ and $Fe^{3+}$ ions [1]. At the Verwey temperature $Tv\sim120K$, conduction electrons freeze at the octahedrally coordinated iron sites forming a charge ordered state. The exact arrangement of the $Fe^{2+}$ ions within the lattice below Tv, and whether short-range or long-range order is present is the subject of a continuing debate. It has even been suggested that the transition may be caused by the dimerisation of electron orbital bonds rather than charge ordering [2].

The surface of magnetite has been studied intensively over the past decade, as it is a highly interesting material for spin electronics. It is one of the few materials with a very high degree of spin polarization at the Fermi level. It is even possible that magnetite is a half metallic ferromagnet, meaning that it is an insulator in one of the two spin bands and a conductor in the other. Spin-electronics almost invariably involves electron transport across or along a material interface. Therefore, an understanding of the structure and electronic properties of surfaces/interfaces is a key issue for making progress in this area. At the same time, progress into exploring the potential of magnetite for spin electronics is slowed by the complexity of the magnetite structure. Magnetite has six possible bulk terminations in the [111] direction [3], but only one of them has been clearly resolved experimentally. The surface is highly sensitive to the preparation conditions. Moreover, self-consistent band structure calculations of magnetite surfaces and surface reconstruction models for low index terminations are still not available.

Apart from possible transformations associated with the rearrangement of atoms at the surface, that should be common for the vast family of spinel oxides, magnetite has an additional unique complexity caused by possible charge rearrangement at the cationic sites. It has been reported that under the correct preparation conditions, the magnetite (100) surface may undergo a metal-insulator transition at a temperature above room temperature [4,5]. It is now clear that the charge order on the (100) surface critically depends on the surface stoichiometry.

Charge ordering on other low index terminations of magnetite has not been studied. However, one can expect that this should also be sensitive to surface stoichiometry. In this paper, we focus on the structural transformations of a $Fe_3O_4$ (111) single crystal surface and their impact on the electron structure and charge ordering at the surface and in the subsurface layer.

## 2. EXPERIMENT

The ultra-high vacuum (UHV) experimental set-up includes a room-temperature scanning tunnelling microscope (STM), a reverse view four-grid Low-Energy Electron Diffraction (LEED) optics, and a cylindrical mirror analyser for Auger electron spectroscopy (AES). The experimental set-up is described in detail in [6].

The magnetite sample was grown by the floating zone technique [7] and cut along the (111) plane with a precision of ±1°. A characterization of the sample by X-ray powder diffractometry and single crystal high-resolution diffractometry using a Cu target ($\lambda$ =1.54 Å) showed good agreement with the reference peaks for magnetite. By means of high resolution X-ray diffractometry we have also verified that the sample miscut was consistent with the expected precision of ±1°. A Verwey transition temperature of 120 K was measured from the discontinuity in the resistivity versus temperature curve, indicating that the crystal is stoichiometric. Measurements of the magnetization of the sample around Tv were performed in fields of 5 to 50 mT, using a vibrating sample magnetometer. The value of saturation magnetization of 94 emu/g obtained from the magnetization measurements at 297 K is in good agreement with the expected value for magnetite. The values of Tv obtained from the magnetization and resistivity measurements are consistent with stoichiometric magnetite. By characterizing the sample using Raman spectroscopy measurements, we were able to positively exclude the presence of other iron oxide phases in the crystal.

The crystal was mechanically polished with diamond paste down to a grain size of 0.05 μm, cleaned using organic solvents and transferred into the UHV system. The in-vacuum sample cleaning procedure consisted of repeated cycles of Ar+ ion etching (1.0-0.5 keV, 5-15 min) and annealing in UHV at a temperature of 950K for a few hours. After a few such cycles, the sample reveals a contaminant free surface, i.e. AES measurements indicate that the concentration of contaminants is below 1% and LEED patterns taken on the surface correspond to an unreconstructed bulk termination. The final *in-situ* surface preparation is crucially important to the results presented in this paper and therefore is described in detail separately for each case.

## 3. RESULTS AND DISCUSSION

The crystal structure of magnetite allows for six ideal bulk terminations of the (111) surface (Fig.1). Two of them represent terminations consisting of close-packed oxygen layers and the other four are surface planes containing cations ($Fe^{2+}$, $Fe^{3+}$). There are significant differences between the four Fe-terminations. One of them consists of Fe tetrahedral sites arranged in a hexagonal (honeycomb) lattice with a 6 Å periodicity (Fig.1c). Another one has octahedral site Fe atoms arranged in a "Kagome" lattice (Fig.1b) with 3 Å interatomic distances. The two remaining Fe-containing terminations are multilayer terminations with octahedral and tetrahedral iron layers separated by 0.6 Å in the [111] direction. One of the two oxygen terminations appears as an oxygen monolayer on top of octahedral Fe, while the other one covers a multilayer of tetrahedral and octahedral Fe atoms. The atomic periodicities of both oxygen terminations are identical (Fig.1d). The bulk separation between nearest neighbour anions in these terminations varies between 2.851 Å and 3.084 Å.

All STM, AES and LEED measurements presented in this publication were performed at room temperature. Typically the STM was operated in the constant-current mode. In our study of the magnetite (111) surface we employ tips that we have fabricated from a variety of materials [8]: nonmagnetic tungsten tips, ferromagnetic Ni tips and also tips made of antiferromagnetic MnNi alloy. Antiferromagnetic tips are interesting for spin polarised STM imaging as first pointed out in ref. [9]. Although this particular study was not aimed at achieving spin-polarised contrast, we employed MnNi tips, as in our experience they also more readily yield conventional atomic resolution images. The results of the spin polarised contrast studies on the $Fe_3O_4$ (111) will be published elsewhere. Details of the scanning conditions are presented with reference to the relevant figures.

We have employed three *in-situ* sample preparation procedures. The initial step of all the three procedures consists of annealing the sample in UHV (base pressure $10^{-11}$ mbar) to a temperature of about 1000K. This treatment has a significant impact on the surface stoichiometry resulting in a reduction of the O/Fe ratio as observed by

AES. The $O_{(510\ eV)}/Fe_{(703\ eV)}$ peak ratio drops to a value of 1.1 after anneal (compared with 1.33 for the stoichiometric surface). Similar results have been recorded for the (100) $Fe_3O_4$ surface [4]. The final anneal step aimed at recovering the subsurface stoichiometry was different for the three *in-situ* preparation procedures. Depending on the details of the final anneal step, this leads to the formation of the so-called "regular" termination, "oxygen" termination or the "intermediate" case. These are described below in detail along with the conditions of the final anneal step.

A. "Regular" termination

This termination was formed as a result of annealing in an oxygen atmosphere of $10^{-6}$ mbar at 950K for 15 min. The oxygen was then promptly pumped out of the chamber and the sample was cooled down to room temperature in UHV conditions. Following this sample treatment, AES measurements indicate an $O_{(510eV)}/Fe_{(703eV)}$ ratio of about 1.3, which is consistent with the value for the magnetite bulk [4]. The LEED analysis of the surface after such annealing yielded patterns with hexagonal symmetry and a periodicity of 6 Å, which is consistent with that of the $Fe_3O_4$ bulk termination (Fig.2a).

The $Fe_3O_4$ (111) Fe-tetrahedral termination is referred to here as "regular", as it is commonly reported for the magnetite [111] surface [12-14]. Fig.3 shows a typical STM image of the "regular" termination. It represents a p(1×1) hexagonal lattice with a 6 Å periodicity. The hexagonal pattern of bright spots corresponds to $Fe^{3+}$ atoms in tetrahedral positions and therefore the "regular" termination can be considered as one of the six possible bulk terminations (Fig.1c). A height difference of about 4.8 Å between terraces was observed by us on this surface, which is also consistent with the model of the tetrahedrally coordinated Fe-cation termination (Fig1a).

B. "Oxygen" termination

The second *in-situ* preparation procedure was performed with the aim of obtaining an oxygen-rich surface of magnetite (111). According to the phase diagram of the Fe-O system [14], iron oxide that is annealed to equilibrium at an oxygen pressure of $10^{-6}$ mbar at a temperature above 900K, is magnetite. Oxygen-rich iron oxide phases, like

α-Fe$_2$O$_3$, start to form at the same oxygen pressure if the annealing temperature is reduced, and the FeO phase is formed at much higher temperature or lower oxygen pressure respectively. Therefore, in order to obtain the oxygen-rich surface we have chosen a second *in-situ* preparation procedure consisting of annealing the sample at 950K in an oxygen partial pressure of 10$^{-6}$ mbar with subsequent cooling to room temperature at this oxygen pressure. We anticipated that this procedure should significantly change the oxygen concentration on the surface. However, we expected that the stoichiometry change would only affect a shallow subsurface layer of the crystal due to the relatively short cooling time.

Auger measurements indicate that following such an anneal, the O$_{(510eV)}$/Fe$_{(703eV)}$ ratio increases to 1.45. The LEED pattern displays fractional order spots around the integral-order spots, due to a well-defined hexagonal superlattice with a periodicity of 42±3 Å (Fig.2b). Fig.4a shows an STM image of the oxygen-terminated surface formed after the second *in-situ* preparation procedure. The surface is covered by superperiodic structures with a 42 Å periodicity, in agreement with the LEED diffraction pattern. These superstructures are highly regular and cover virtually the entire sample surface. Higher resolution STM images (Fig.4b) show the atomic arrangement within these superperiodic structures. The pattern is made by the repetition of three distinct areas, marked as areas I, II and III in Fig.4b. Detailed analysis shows that area I has a periodicity of 3.1±0.1 Å, while areas II and III both have a periodicity of 2.8±0.1 Å along the [011] direction. The atomic arrangements in areas II and III have exactly the same symmetry and periodicity. The LEED pattern is well in line with the STM results. The separation between the satellite spots and the integral-order spots is consistent with a 42Å superperiodicity, whereas the positions of the doubled main spots are consistent with the presence of two periodicities: 2.8Å and 3.1Å. (Fig.2b).

Superstructures formed on the Fe$_3$O$_4$ (111) surface were reported previously in ref. [10,11] and interpreted as a biphase structure consisting of two phases: FeO and Fe$_3$O$_4$, arranged in a superlattice. This interpretation cannot be applied to our results for a number of reasons. First, the LEED pattern (Fig.2b) does not suggest formation of another iron oxide phase like FeO (the LEED pattern should show floreting about the first and second zone spots in the case of FeO and $Fe_3O_4$ phases simultaneously present on the surface [11]). Second, we expect that following our in-situ preparation procedure the surface is oxidised rather than reduced and this is indeed confirmed by the increase in the O/Fe AES ratio. Therefore, the formation of a reduced iron oxide, FeO, is not expected in our experiment. Concerning of the possibility of forming higher oxidation iron oxide phase on magnetite surface, we should note here that the structural difference between $Fe_3O_4$ and $Fe_2O_3$ [1] makes it easy to distinguish on the LEED pattern. Similar superstructure formed on α–$Fe_2O_3$ surface was reported in [15] and explained as a biphase structure of α–$Fe_2O_3$ (0001) and FeO (111). Compare the LEED pattern in Fig.2 and in [15] we could state no indications of the $Fe_2O_3$ (0001) phase on the surface in our case.

It would be in line with our sample preparation procedure and Auger results to suggest, that the 3.1 Å and 2.8 Å periodicities of the areas I, II and III in the "oxygen" termination result from a closed-packed oxygen layer formed on the magnetite surface. There are additional experimental evidences in favour of the oxygen-termination model. First, in the "oxygen" termination, the atomically resolved STM images could be obtained much more readily at a negative sample bias, suggesting that we likely image the filled states of electronegative surface ions. Second, the smallest separation step between terraces is 2.5±0.6 Å, which is consistent with the distance between two nearest neighbour oxygen layers in the magnetite [111] bulk structure.

Another interesting property of the observed superstructure is that the boundaries between the areas I, II and III shift significantly depending on the bias voltage applied to the sample, resulting in changes of the shape of the superstructure. This is shown in Fig.4c. It is interesting to note that the shape of the superstructure was also altered when different tips (W, Ni, MnNi) were used. The superstructure periodicity does not depend on the bias and the tip material used. However, we should point out that the height difference between the areas I, II and III does depend on the tunnel bias. These facts evidently indicate that the superstructure depicts an electronic rather than a structural effect.

We were able to reproducibly toggle between the "regular" and "oxygen" terminations. For example, we could readily switch the surface to the "regular" termination by subjecting the sample with "oxygen" termination to a 15-minute anneal as described above. Alternatively, we could remove the subsurface layer by $Ar^+$ ion sputtering (10min, 1keV) and again form the "oxygen" termination with the superstructure by repeating the oxidation procedure.

The "oxygen" terminated magnetite surface was found to be very stable. Two days after the sample preparation, we were still able to achieve atomic resolution STM on the oxygen-terminated surface provided the sample was during that time kept under UHV.

C. "Intermediate" case

In order to better understand the formation of the oxygen-termination and, in particular, the superstructure within it, we reduced the temperature of the final anneal step in oxygen to 850 K. The aim was to form a partially closed-packed oxygen layer on the surface and establish how this affects the formation of the superstructure. Following this anneal, the LEED pattern of the surface was similar to the one of the oxygen- termination (Fig.2b), but with weaker satellite spots. Fig.5 shows the STM images corresponding to this altered sample preparation. Three types of terraces are seen in Fig.5: terrace **A** is covered by a superstructure similar to the one described above (oxygen termination); terrace **C** exhibits an hexagonal lattice with a 6 Å periodicity; while another terrace **B** represents the combination of the above two structures. The hexagonal lattice with 6 Å periodicity, however, is not equivalent to the "regular" termination. A zoom into the area $\alpha$ of terrace **B** is shown in Fig.5.3. Most of the 6 Å protrusions split into three atoms positioned with a local triangular geometry and a periodicity of 3 Å. Many protrusions contain only two atoms and therefore correspond to the locations of defects where the oxygen atoms are missing. Some of the latter are indicated with arrows (Fig.5.3). We suggest that terrace **C** and the *α*-region of terrace **B** consist of an incomplete layer of oxygen atoms and is different from the "oxygen" termination described above in that the layer is not complete. The separation between the terraces (2.5±0.6 Å) is essentially identical to the one observed in the case of "oxygen" terminated planes.

The results explained in this section provide further support to our interpretation of the 42 Å superstructure being a long range electronic effect triggered by the change of the O/Fe ratio on the surface. A representative case is shown in Fig.5.2. The upper part (**β**) of the terrace **B** is a segment having a completed oxygen termination, i.e. having a very small number of defects. Such areas invariably form long-range order identical to the one described above. A zoom-in of the region **β** is shown in Fig.5.4. If the oxygen layer contains defects and therefore is incomplete (region **α**), the long-range order is not formed. From our STM results we could estimate that the critical defect density that destroys formation of the long-range order at room temperature is about 10-20%. One could envisage that the long-range order may still be formed on such surfaces with defects at a lower temperature. In other words, the temperature of the long-range order transition is a function of the density of defects.

The 42 Å periodicity of the superstructure was not maintained in the vicinity of the terrace edges and also in the vicinity of the boundaries between areas with the complete oxygen termination and termination with defects (Fig.5.2).

4. MODEL OF THE SUPERSTRUCTURE FORMATION

Our model relies on the existence of a tensile strain (lattice deformation) at the oxygen-terminated surface. The presence of this lattice deformation is evident from our STM data. To further validate the notion of such deformation, we have performed Density Functional Theory (DFT) calculations for the structure of the (111) surface of magnetite. The CASTEP algorithm within the local density approximation of DFT [16] was used to calculate total energies and optimise the atomic positions on the surface to minimise the energy. A good example demonstrating a successful corroboration between STM experiment on transition metal oxides and DFT calculations was provided in recent publication [17].

As explained above, the surface with 3 Å periodicity could represent two possible types of closed-packed oxygen layer termination located either above a layer of B-site $Fe^{2+,3+}$ cations, or an Fe-multilayer with a mixture of A- and B-site cations (Fig.1). The total charge density distribution calculated for both types of oxygen terminations

(Fig.6) show the local charge density maximas above the oxygen sites, which is in agreement with our STM data. We were not able to run a geometry optimisation calculation for an extended slab to find out the condition of the superstructure formation on (111) surface. Nevertheless, the calculation for one unit cell slab indicates that the positions of some oxygen atoms shifted laterally (Fig.6) with respect to their bulk positions and therefore point to the formation of strain along the surface of the oxygen termination. The mechanism of such strain is clear: dangling O-bonds at the surface result in excess electrons that are retained in the topmost oxygen layer. The resulting Coulomb repulsion established between the oxygen anions containing these excess electrons produces an increase in inter-ionic separation from 2.8 Å to 3.1 Å.

As we pointed above the superstructure represents an electronic rather than simply a structural effect, meaning that there is a significant difference in tunnelling conductance in areas I-III. There are two types of instabilities in crystals that link together excessive strain and changes in electronic properties and create a long-range order: localised polaron and Charge Density Waves (CDW). The concept of polarons has been applied to magnetite in the past. For example, it was suggested in [18], that at room temperature the conductivity of magnetite could be explained in terms of polaronic transport. Particularly, the concept of Jahn-Teller polarons as local distortions of the lattice around $Fe^{3+}$ ions was proposed in [18, 19]. Another possibility is to base the model on the concept of a CDW. The CDW is also an electronically induced lattice instability accompanied by a modulation of the electronic density. In the case of one-dimensional systems, a CDW is known to occur at a wave vector $q = 2k_F$ and is known as Peierls instability. In the case of 2D and 3D systems, a CDW can develop for Fermi liquids with a nested Fermi surface [20]. The surface structure of magnetite could be particularly susceptible to modulation of the electronic structure due to the presence of $Fe^{2+}$ and $Fe^{3+}$ ions at the B-sites. The strain could cause re-allocation of charges between the octahedral B sites. Reallocation of charges between A- and B-sites could be also considered.

To complete the model further experiments are needed to explore electronic and magnetic properties of the reported superstructure.

## 5. SUMMARY

Three different terminations on the (111) surface of a synthetic $Fe_3O_4$ single crystal were formed and studied using LEED, AES and STM with a range of different nonmagnetic, ferromagnetic and antiferromagnetic tip materials. The three sample preparation procedures leading to the three terminations varied only in the conditions of the last step: *in-situ* anneal in oxygen. They result in three different values of the surface stoichiometry. The three surface terminations were identified as a termination of Fe cations in tetrahedral sites called the "regular" termination, complete oxygen layer termination called the "oxygen" termination and the intermediate case of an oxygen layer termination containing defects.

The oxygen-terminated surface forms an hexagonal superstrucure with periodicity of 42 Å. In the superstructure, the atomic level periodicities are virtually unaffected except for the modulation of strain that could be observed using STM and LEED. The change of the superstructure shape depends on the bias voltage. It is also sensitive to the type of tip material used in the STM. We conclude that the superstructure is an electronic effect resulting from an electron-lattice instability, rather than a mosaic of different iron oxide phases. The superstructure is sensitive to the oxygen to iron ratio at the surface. Once a nearly complete layer of oxygen covers the surface, lateral strain is developed triggering the formation of the superstructure. On the other hand, if the layer contains defects in the form of missing oxygen atoms, the strain is reduced and the superstructure disappears. From our STM data we could work out that the critical density of defects destroying the formation of the superstructure at room temperature is approximately 10-20% of the complete oxygen-terminated layer. We have performed DFT calculations for the "regular" and two different "oxygen" terminations. The results of the calculations suggest that unlike in the "regular" termination, each of the two possible oxygen terminations results in a significant lateral strain that may lead to the formation of long-range order through electron-lattice instability. We suggest either polaronic or CDW electron-lattice instability may develop the superstructure on the surface of magnetite.

## V. ACKNOWLEDGMENTS

This work was supported by Science Foundation of Ireland (SFI) under contract 00/PI.1/C042.

**Figure Captions**

FIG.1. Magnetite [111] slab (a); two different bulk terminations that contain Fe cations (b,c); and oxygen termination (d). In figure b and c, oxygen layer shown are positioned below the surface termination. Figure d has enlarged scale.

FIG .2 LEED patterns of $Fe_3O_4$ (111) surface at E=67 eV: (a) tetrahedral site Fe termination, 6.0±0.3 Å periodicity indicated, and (b) $Fe_3O_4$ (111) overoxidized surface. Two large rhombus indicate the periodicities of 3.1±0.1 Å and 2.8 ± 0.1 Å; the small rhombus in the insert shows a 42±3 Å superstructure.

FIG. 3. 15×15nm$^2$ STM image of $Fe_3O_4$ (111) showing the "regular" type of termination with Fe-ions in tetrahedral sites with 6.0±0.3 Å periodicity. ($V_{bias}$= 0.3 V, $I_t$ = 0.1 nA, MnNi tip).

Fig. 4. (a) 120×120nm$^2$ STM image of overoxidized $Fe_3O_4$ (111) surface representing a large terraces covered by superstructures of about 42 Å period; (b) 10×8 nm$^2$ STM image of superperiodic pattern seen in (a). Area I has 3.1±0.1 Å interatomic periodicity, and areas II and III have 2.8 ± 0.1 Å periodicity ($V_{bias}$ = -1.0V, $I_t$=0.1nA,MnNi tip). (c) 15×10nm$^2$ STM image of the same sample as in (a), with different bias: $V_{bias}$ = -0.5 V. The height difference between areas II and III now is 0.4 Å, instead of 0.6 Å as in (b) ($V_{bias}$ = -0.5V, $I_t$=0.1nA,MnNi tip).

Fig.5. $Fe_3O_4$ (111) surface after annealing in oxygen at 850K: 30x30nm$^2$ (1) and 20x20nm$^2$ (2) STM images showing few terraces with different atomic arrangement: terrace (**A**) completely covered by superstructure; terrace (**C**) represents region of 6 Å periodicity; terrace (**B**) has a mixture of atomic arrangements presented on terraces **A** and **C**; (3) zoom-in on region **α**, 6 Å periodicity region showing the splitting to 3 Å arrangement (arrows indicate the oxygen defects); (4) zoom-in of superstructure region **β**. ($V_{bias}$= -1.0 V, $I_t$ = 0.1 nA, Ni tip).

FIG.6. $Fe_3O_4$ (111) crystal structure: Fe-tetrahedral termination (left-top); oxygen termination (left-bottom) with the 0.2 level of electron density is shown as iso-surface; 2D-map of the total charge density (right) above surface calculated for the vacuum slabs shown on the left. Hexagonal symmetry of 3 Å oxygen lattice is distorted by strain as shown by triangles marking atom positions. DFT calculations were performed for one unit cell slab.

**Fig.1.** Magnetite [111] slab (a); two different bulk terminations that contain Fe cations (b,c); and oxygen termination (d). In figure b and c, oxygen layer shown are positioned below the surface termination. Figure d has enlarged scale.

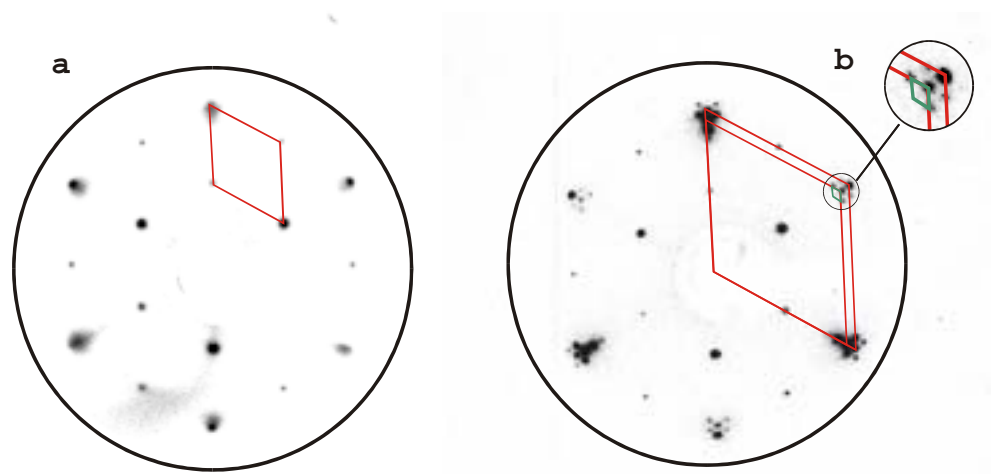

**Fig.2.** LEED patterns of Fe$_3$O$_4$ (111) surface at E=67 eV. (a) tetrahedral site Fe termination, 6.0±0.3 Å periodicity indicated, and (b) Fe$_3$O$_4$ (111) overoxidized surface. Two large rhombs indicate the periodicities of 3.1±0.1 Å and 2.8 ± 0.1 Å; the small rhombus at the highlight shows 42±3 Å superstructure.

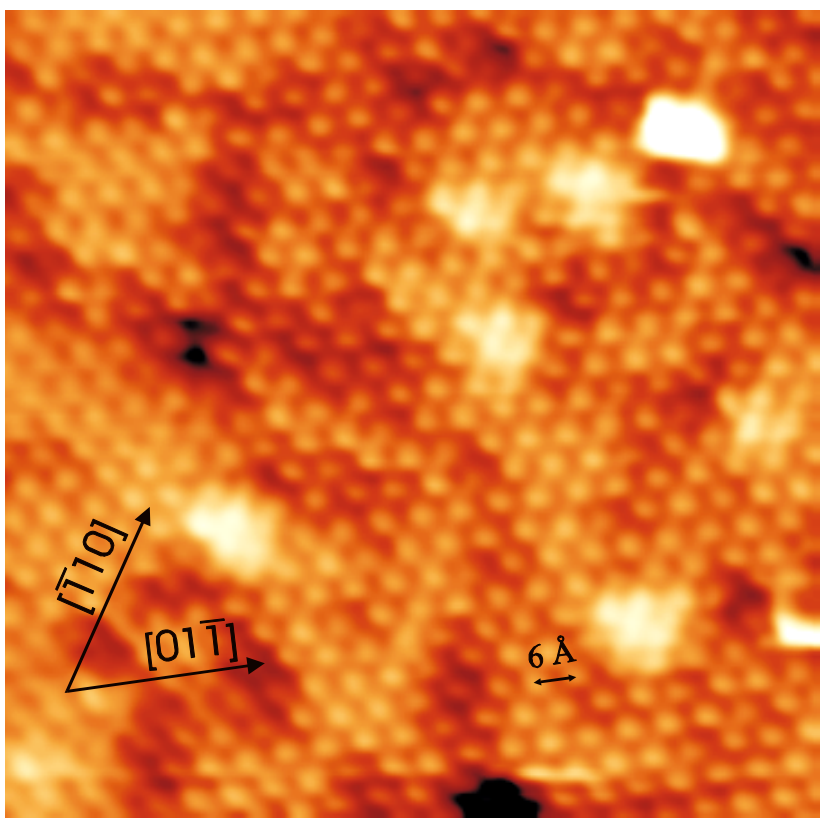

**Fig. 3**. 15×15nm² STM image of $Fe_3O_4$ (111) showing the "regular" type of termination with Fe in tetrahedral sites of 6.0±0.3 Å periodicity. ($V_{bias}$= 0.3 V, $I_t$ = 0.1 nA, MnNi tip).

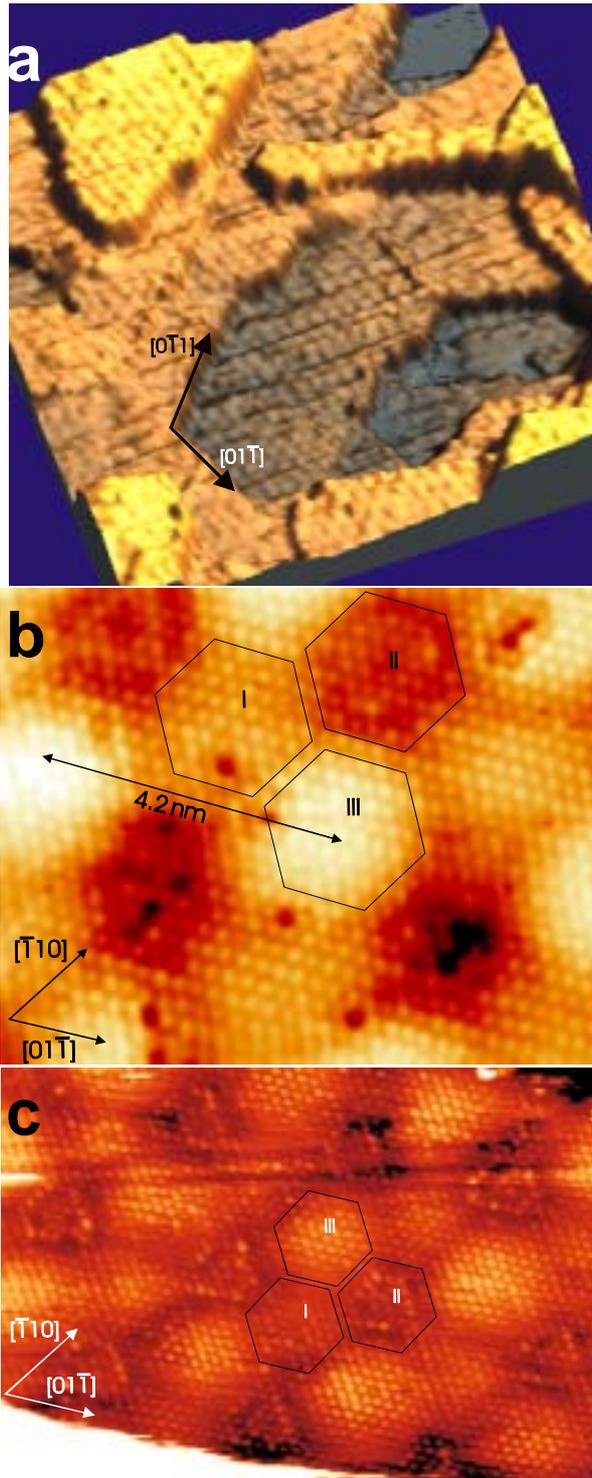

Fig. 4. (a) 120×120nm$^2$ STM image of overoxidized $Fe_3O_4$ (111) surface representing a large terraces covered by superstructures of about 42 Å period (b) 10×8 nm$^2$ STM image of superperiodic pattern seen in (a). Area I has 3.1±0.1 Å interatomic periodicity, and areas II and III have 2.8 ± 0.1 Å periodicity ($V_{bias}$ = -1.0V, $I_t$=0.1nA, MnNi tip). (c) 15×10nm$^2$ STM image of the same sample as in (a), with different bias: $V_{bias}$ = -0.5 V. The height difference between areas II and III now is 0.4 Å, instead of 0.6 Å as in (b) ($V_{bias}$ = -0.5V, $I_t$=0.1nA, MnNi tip).

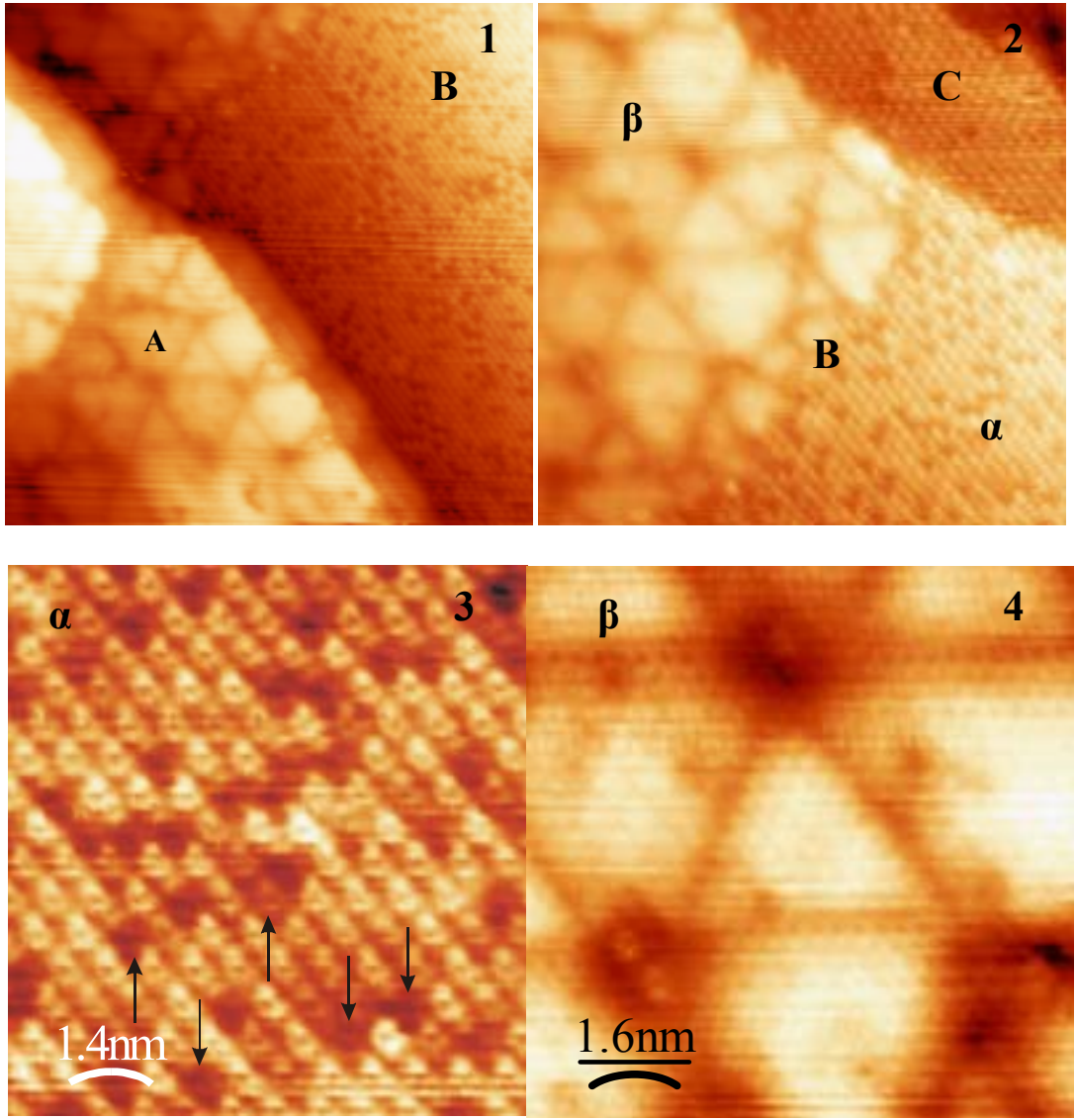

Fig.5. Fe$_3$O$_4$ (111) surface after annealing in oxygen at 850K: 30x30nm$^2$ (1) and 20x20nm$^2$ (2) STM images showing few terraces with different atomic arrangement: terrace (**A**) completely covered by superstructure; terrace (**C**) represents region of 6 Å periodicity; terrace (**B**) has a mixture of atomic arrangements presented on terraces **A** and **C**; (3) zoom-in on region **α**, 6 Å periodicity region showing the splitting to 3 Å arrangement (arrows indicate the oxygen defects); (4) zoom-in of superstructure region **β**. (V$_{bias}$= -1.0 V, I$_t$ = 0.1 nA, Ni tip).

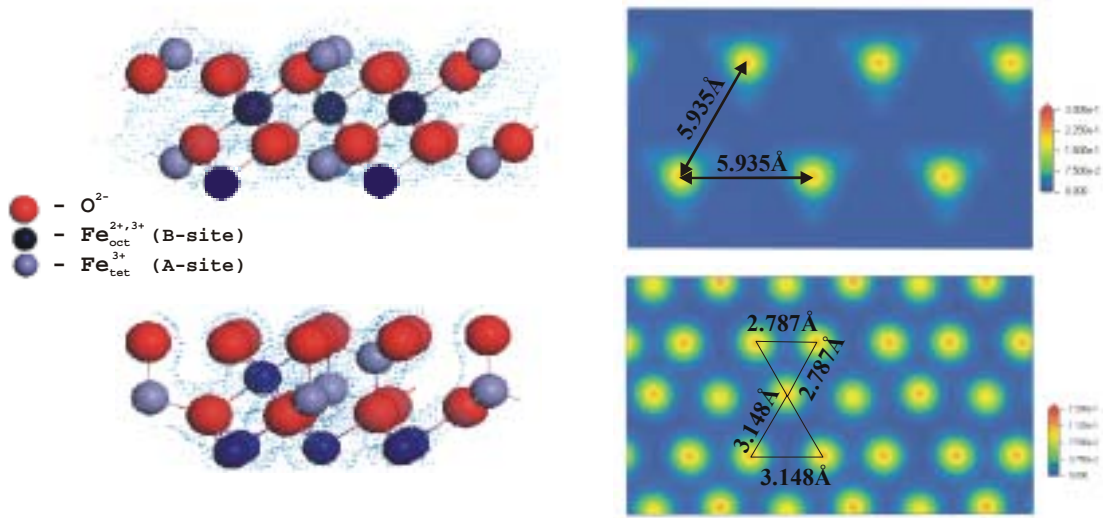

**Figure 6.** $Fe_3O_4$ (111) crystal structure: Fe-tetrahedral termination (left-top); oxygen termination (left-bottom) with the 0.2 level of electron density is shown as iso-surface; 2D-map of the total charge density (right) above surface calculated for the vacuum slabs shown on the left. Hexagonal symmetry of 3 Å periodic structure is distorted by stress as shown by triangles marking atom positions. Geometry optimization is performed for one unit cell slab.